# A family of asymptotically good quantum codes based on code concatenation

Zhuo Li,[*]  Li-Juan Xing,[†]  and Xin-Mei Wang

State Key Laboratory of Integrated Service Networks, Xidian University, Xi'an,

Shannxi 710071, China

We explicitly construct an infinite family of asymptotically good concatenated quantum stabilizer codes where the outer code uses CSS-type quantum Reed-Solomon code and the inner code uses a set of special quantum codes. In the field of quantum error-correcting codes, this is the first time that a family of asymptotically good quantum codes is derived from bad codes. Its occurrence supplies a gap in quantum coding theory.

Quantum error correction is a basic technique for transmitting quantum information reliably over a noisy quantum channel. Many explicit constructions of quantum error-correcting codes have been proposed so far [1-12]. Some of the best-known code constructions are the CSS code construction of Calderbank and Shor [1] and Steane [2] and the stabilizer code construction of Gottesman [3] and Calderbank et al. [4, 5].

As in classical coding theory, we want to construct quantum codes with large minimum distance. More generally, we want to construct asymptotically good quantum codes with both rate and distance/length bounded away from zero. Ashikhmin et al. [13] and Chen et al. [14] constructed asymptotically good quantum codes based on algebraic geometry codes. Later, Matsumoto [15] improved the bound of Ashikhmin et al. [13].

In classical coding theory, code concatenation [16] is a basic method for constructing good error-correcting codes and most of the known asymptotically good binary codes are constructed by code concatenation [17]. In the quantum setting, code concatenation is also effectively used to construct good quantum error-correcting codes, although concatenation is mainly used for fault-tolerant quantum computation [18]. Gottesman states code concatenation in his PhD thesis and gives the stabilizer of a quantum code constructed by concatenating the five-qubit code with itself. Calderbank et al. [5] also remark concatenated codes and Rains [19] proves the so-called product bound of concatenated codes.

In this paper we derive an infinite family of asymptotically good binary quantum stabilizer codes from quantum Reed-Solomon (RS) codes, which may be thought of as concatenated quantum codes where the outer code is CSS-type quantum RS code and the inner codes use $N$ distinct quantum codes. Firstly we give the structures of the stabilizer and normalizer. Then we show that this family of codes is asymptotically good. These codes are distinguished by being the first family of codes we have seen with the property that good quantum codes are obtained from bad codes. For this we first show that the long binary quantum codes obtained from RS codes are bad briefly. Let $\mathcal{C}$ be an $[\![mN, m(N-2K), K+1]\!]$ binary quantum code obtained from RS code [20]. If the rate $= m(N-2K)/mN = (N-2K)/N$ is held fixed, the ratio distance/length $=(K+1)/mN$ approaches zero as $m \to \infty$. However, by a very clever construction it is possible to obtain an infinite family of good binary quantum codes from RS codes, as we now show.

***Lemma 1 [12].*** (CSS codes) Let $C$ be an $[n,k]_q$ weakly self-dual code over $GF(q)$ and let $d = \min\{wgt(v) : v \in C^\perp \setminus C\}$. Then there exists a quantum stabilizer code encoding $n-2k$ qudits into $n$ qudits with minimum distance $d$, denoted by $[\![n, n-2k, d]\!]_q$ with stabilizer $\mathcal{S} = C \times C$ and normalizer $\mathcal{N} = C^\perp \times C^\perp$.

The starting point is a CSS-type quantum RS code $[\![N, N-2K, K+1]\!]_{2^{2m}}$ with stabilizer $\mathcal{S}_{RS} = \mathcal{R} \times \mathcal{R}$ and normalizer $\mathcal{N}_{RS} = \mathcal{R}^\perp \times \mathcal{R}^\perp$ where $\mathcal{R} = [N, K, N-K+1]_{2^{2m}}$ and $\mathcal{R}^\perp = [N, N-K, K+1]_{2^{2m}}$ are classical RS codes over $GF(2^{2m})$ with $\mathcal{R} \subseteq \mathcal{R}^\perp$ for $N = 2^{2m}-1$ and $K \leq \lfloor N/2 \rfloor$. Let $\alpha$ be a primitive element of $GF(2^{2m})$ and let $\beta_1, \ldots, \beta_{2m}$ be a self-dual basis of $GF(2^{2m})$ over $GF(2)$. Let $a = (a_0, \ldots, a_{N-1} | a_N, \ldots, a_{2N-1})$, $a_i \in GF(2^{2m})$, be a typical codeword of $\mathcal{S}_{RS}$ with $a_i = \sum_{j=1}^{2m} a_{i,j} \beta_j$, $a_{i,j} \in GF(2)$. Let $b$ be the binary vector

$$b = (b_{0,1}, \ldots, b_{0,4m+2}; \ldots; b_{N-1,1}, \ldots, b_{N-1,4m+2} | c_{0,1}, \ldots, c_{0,4m+2}; \ldots; c_{N-1,1}, \ldots, c_{N-1,4m+2})$$

such that

$$\sum_{j=1}^{2m} b_{i,j} \beta_j = \alpha^{-i}[(a_{N+i,1} + s_{i,m+1})\beta_1 + \sum_{j=2}^{m} a_{N+i,j} \beta_j + \sum_{j=m+1}^{2m} s_{i,j-m} \beta_j], \quad b_{i,2m+1} = a_{i,1} + s_{i,1},$$

$$\sum_{j=1}^{2m} b_{i,2m+1+j} \beta_j = \alpha^{-i}[(a_{N+i,m+1} + t_{i,m+1})\beta_1 + \sum_{j=2}^{m} a_{N+i,m+j} \beta_j + \sum_{j=m+1}^{2m} t_{i,j-m} \beta_j], \quad b_{i,4m+2} = a_{i,m+1} + t_{i,1},$$

$$\sum_{j=1}^{2m} c_{i,j} \beta_j = \alpha^{i}(\sum_{j=1}^{m} a_{i,j} \beta_j + s_{i,m+1} \beta_{m+1}), \quad c_{i,2m+1} = s_{i,m+1},$$

$$\sum_{j=1}^{2m} c_{i,2m+1+j} \beta_j = \alpha^i (\sum_{j=1}^{m} a_{i,m+j} \beta_j + t_{i,m+1} \beta_{m+1}), \quad c_{i,4m+2} = t_{i,m+1},$$

where $s_{i,j}, t_{i,j} \in GF(2)$ for $0 \le i \le N-1$, $1 \le j \le m+1$. Let

$$a^* = (a_0^*, \ldots, a_{N-1}^* | a_N^*, \ldots, a_{2N-1}^*), \quad a_i^* \in GF(2^{2m}),$$ be a typical codeword of $\mathcal{N}_{RS}$ with

$$a_i^* = \sum_{j=1}^{2m} a_{i,j}^* \beta_j, \quad a_{i,j}^* \in GF(2).$$ Let $b^*$ be the binary vector

$$b^* = (b_{0,1}^*, \ldots, b_{0,4m+2}^*; \ldots; b_{N-1,1}^*, \ldots, b_{N-1,4m+2}^* | c_{0,1}^*, \ldots, c_{0,4m+2}^*; \ldots; c_{N-1,1}^*, \ldots, c_{N-1,4m+2}^*)$$

such that

$$\sum_{j=1}^{2m} b_{i,j}^* \beta_j = \alpha^{-i}[(a_{N+i,1}^* + s_{i,m+1}^*)\beta_1 + \sum_{j=2}^{m} a_{N+i,j}^* \beta_j + \sum_{j=m+1}^{2m} s_{i,j-m}^* \beta_j], \quad b_{i,2m+1}^* = a_{i,1}^* + s_{i,1}^*,$$

$$\sum_{j=1}^{2m} b_{i,2m+1+j}^* \beta_j = \alpha^{-i}[(a_{N+i,m+1}^* + t_{i,m+1}^*)\beta_1 + \sum_{j=2}^{m} a_{N+i,m+j}^* \beta_j + \sum_{j=m+1}^{2m} t_{i,j-m}^* \beta_j], \quad b_{i,4m+2}^* = a_{i,m+1}^* + t_{i,1}^*,$$

$$\sum_{j=1}^{2m} c_{i,j}^* \beta_j = \alpha^i (\sum_{j=1}^{m} a_{i,j}^* \beta_j + s_{i,m+1}^* \beta_{m+1}), \quad c_{i,2m+1}^* = s_{i,m+1}^*,$$

$$\sum_{j=1}^{2m} c_{i,2m+1+j}^* \beta_j = \alpha^i (\sum_{j=1}^{m} a_{i,m+j}^* \beta_j + t_{i,m+1}^* \beta_{m+1}), \quad c_{i,4m+2}^* = t_{i,m+1}^*,$$

where $s_{i,j}^*, t_{i,j}^* \in GF(2)$ for $0 \le i \le N-1$, $1 \le j \le m+1$. Let $\mathcal{S}_\mathcal{L}$ denote the code consisting of all such vectors $b$ and let $\mathcal{N}_\mathcal{L}$ denote the code consisting of all such vectors $b^*$.

***Lemma 2.*** $\mathcal{S}_\mathcal{L}$ is dual to $\mathcal{N}_\mathcal{L}$ with respect to the symplectic inner product.

***Proof.*** Clearly $\mathcal{S}_\mathcal{L}$ and $\mathcal{N}_\mathcal{L}$ are binary linear codes. From definitions of $b$ and $b^*$ we have

$$\sum_{j=1}^{2m}(b_{i,j} c_{i,j}^* + b_{i,j}^* c_{i,j}) = \text{Tr}\left( (\sum_{j=1}^{2m} b_{i,j} \beta_j)(\sum_{j=1}^{2m} c_{i,j}^* \beta_j) + (\sum_{j=1}^{2m} c_{i,j} \beta_j)(\sum_{j=1}^{2m} b_{i,j}^* \beta_j) \right)$$

$$= \text{Tr}\left( [(a_{N+i,1} + s_{i,m+1})\beta_1 + \sum_{j=2}^{m} a_{N+i,j} \beta_j + \sum_{j=m+1}^{2m} s_{i,j-m} \beta_j](\sum_{j=1}^{m} a_{i,j}^* \beta_j + s_{i,m+1}^* \beta_{m+1}) \right.$$

$$+[(a^*_{N+i,1}+s^*_{i,m+1})\beta_1+\sum_{j=2}^{m}a^*_{N+i,j}\beta_j+\sum_{j=m+1}^{2m}s^*_{i,j-m}\beta_j](\sum_{j=1}^{m}a_{i,j}\beta_j+s_{i,m+1}\beta_{m+1}))$$

$$=\sum_{j=1}^{m}(a_{N+i,j}a^*_{i,j}+a^*_{N+i,j}a_{i,j})+a^*_{i,1}s_{i,m+1}+a_{i,1}s^*_{i,m+1}+s_{i,1}s^*_{i,m+1}+s^*_{i,1}s_{i,m+1},$$

$$\sum_{j=2m+2}^{4m+1}(b_{i,j}c^*_{i,j}+b^*_{i,j}c_{i,j})=\sum_{j=1}^{m}(a_{N+i,m+j}a^*_{i,m+j}+a^*_{N+i,m+j}a_{i,m+j})$$

$$+a^*_{i,m+1}t_{i,m+1}+a_{i,m+1}t^*_{i,m+1}+t_{i,1}t^*_{i,m+1}+t^*_{i,1}t_{i,m+1}.$$

Thus the symplectic inner product

$$\sum_{i=0}^{N-1}\sum_{j=1}^{4m+2}(b_{i,j}c^*_{i,j}+b^*_{i,j}c_{i,j})=\sum_{i=0}^{N-1}[\sum_{j=1}^{2m}(b_{i,j}c^*_{i,j}+b^*_{i,j}c_{i,j})+(b_{i,2m+1}c^*_{i,2m+1}+b^*_{i,2m+1}c_{i,2m+1})$$

$$+\sum_{j=2m+2}^{4m+1}(b_{i,j}c^*_{i,j}+b^*_{i,j}c_{i,j})+(b_{i,4m+2}c^*_{i,4m+2}+b^*_{i,4m+2}c_{i,4m+2})]$$

$$=\sum_{i=0}^{N-1}\sum_{j=1}^{2m}(a_{i,j}a^*_{N+i,j}+a^*_{i,j}a_{N+i,j})=0.$$

Then the statement follows by a dimension argument. Q.E.D.

Clearly $\mathcal{N}_\mathcal{L}$ contains $\mathcal{S}_\mathcal{L}$. Thus $\mathcal{S}_\mathcal{L}$ is weakly self-dual under symplectic inner product by Lemma 2. Then a quantum stabilizer code can be derived from $\mathcal{S}_\mathcal{L}$.

**Definition.** For any $N$ and $K$, define $\mathcal{L}_{N,K}$ to be the quantum stabilizer code with stabilizer $\mathcal{S}_\mathcal{L}$ and normalizer $\mathcal{N}_\mathcal{L}$ which are obtained from the CSS-type quantum RS code $[\![N,N-2K,K+1]\!]_{2^{2m}}$.

Clearly $\mathcal{L}_{N,K}$ is a binary quantum stabilizer code with parameters $[\![2N(2m+1),2m(N-2K)]\!]_2$. In the other hand, $\mathcal{L}_{N,K}$ may be thought of as concatenated quantum code where the outer code is CSS-type quantum RS code and the inner codes use $N$ distinct quantum codes.

Before proving the first theorem we need some lemmas. These involve the entropy function $H_4(x)$, defined by

$$H_4(x) = -x \log_4 \frac{x}{3} - (1-x) \log_4 (1-x)$$

where $0 \leq x \leq 1$. We shall also need the inverse function $H_4^{-1}(y)$ defined by

$$x = H_4^{-1}(y) \text{ iff } y = H_4(x)$$

for $0 \leq x \leq \frac{3}{4}$.

***Lemma 3.*** Suppose $\lambda n$ is an integer, where $0 < \lambda < \frac{3}{4}$. Then

$$\sum_{k=0}^{\lambda n} 3^k \binom{n}{k} \leq 4^{nH_4(\lambda)}.$$

***Proof.*** For any negative number $r$ we have

$$4^{r\lambda n} \sum_{k=0}^{\lambda n} 3^k \binom{n}{k} \leq \sum_{k=0}^{\lambda n} 4^{rk} 3^k \binom{n}{k} \leq \sum_{k=0}^{n} (3 \cdot 4^r)^k \binom{n}{k} = (1 + 3 \cdot 4^r)^n.$$

Thus

$$\sum_{k=0}^{\lambda n} 3^k \binom{n}{k} \leq \{4^{-r\lambda} + 3 \cdot 4^{r(1-\lambda)}\}^n.$$

Choose $r = \log_4(\lambda/3(1-\lambda))$. Then this sum is

$$\leq 4^{nH_4(\lambda)} (1 - \lambda + \lambda)^n = 4^{nH_4(\lambda)}. \qquad \text{Q.E.D.}$$

***Lemma 4.*** If we are given $M$ distinct nonzero quaternary $L$-tuples (vectors of length $L$ over $GF(4)$), where

$$M = \gamma(4^{\delta L} - 1), \quad 0 < \gamma < 1, \quad 0 < \delta < 1,$$

then the sum of the weights of these $L$-tuples is at least

$$L\gamma(4^{\delta L} - 1)(H_4^{-1}(\delta) - o(L)).$$

***Proof.*** The number of these $L$-tuples having weight $\leq \lambda L$ is at most

$$\sum_{i=1}^{\lambda L} 3^k \binom{L}{i} \leq 4^{LH_4(\lambda)},$$

by Lemma 3, for any $0 < \lambda < \frac{3}{4}$.

So the total weight is at least

$$\lambda L(M - 4^{LH_4(\lambda)}) = \lambda LM(1 - 4^{LH_4(\lambda)}/M).$$

Choose $\lambda = H_4^{-1}(\delta - 1/\log L) = H_4^{-1}(\delta) - o(L)$, with $\lambda < \frac{3}{4}$. Then the total weight is at least

$$L\gamma(4^{\delta L} - 1)(1 - o(L))(H_4^{-1}(\delta) - o(L)),$$

$$= L\gamma(4^{\delta L} - 1)(H_4^{-1}(\delta) - o(L)). \qquad \text{Q.E.D.}$$

***Theorem 5.*** Let $R$ be fixed, $0 < R < \frac{1}{2}$. For each $m$ choose

$$K = \left\lfloor \frac{1}{2}(1 - \frac{2m+1}{m}R)N \right\rfloor = \left\lfloor \frac{1}{2}(1 - \frac{2m+1}{m}R)(2^{2m} - 1) \right\rfloor.$$

Then $\mathcal{L}_{N,K}$ is a binary stabilizer code of length $n = 2N(2m+1)$ with rate

$$R_m = \frac{m(N - 2K)}{N(2m+1)} \geq R,$$

and a lower bound on distance/length equal to

$$\frac{H_4^{-1}(1/4)}{4}(1 - 2R) \quad \text{as} \quad m \to \infty.$$

***Proof.*** Let $c = (b_{0,1}, \ldots, b_{0,4m+2}; \ldots; b_{N-1,1}, \ldots, b_{N-1,4m+2} \mid c_{0,1}, \ldots, c_{0,4m+2}; \ldots; c_{N-1,1}, \ldots, c_{N-1,4m+2})$

be any nonzero codeword of $\mathcal{N}_\mathcal{L} \setminus \mathcal{S}_\mathcal{L}$. As we saw earlier, there must exists a nonzero codeword $a = (a_0, \ldots, a_{N-1} \mid a_N, \ldots, a_{2N-1}) \in \mathcal{N}_{RS}$ with $a_i = \sum_{j=1}^{2m} a_{i,j} \beta_j$, such

that

$$\sum_{j=1}^{2m} b_{i,j}\beta_j = \alpha^{-i}[(a_{N+i,1} + s_{i,m+1})\beta_1 + \sum_{j=2}^{m} a_{N+i,j}\beta_j + \sum_{j=m+1}^{2m} s_{i,j-m}\beta_j], \quad b_{i,2m+1} = a_{i,1} + s_{i,1},$$

$$\sum_{j=1}^{2m} b_{i,2m+1+j}\beta_j = \alpha^{-i}[(a_{N+i,m+1}+t_{i,m+1})\beta_1 + \sum_{j=2}^{m} a_{N+i,m+j}\beta_j + \sum_{j=m+1}^{2m} t_{i,j-m}\beta_j], \quad b_{i,4m+2} = a_{i,m+1}+t_{i,1},$$

$$\sum_{j=1}^{2m} c_{i,j}\beta_j = \alpha^i (\sum_{j=1}^{m} a_{i,j}\beta_j + s_{i,m+1}\beta_{m+1}), \quad c_{i,2m+1} = s_{i,m+1},$$

$$\sum_{j=1}^{2m} c_{i,2m+1+j}\beta_j = \alpha^i (\sum_{j=1}^{m} a_{i,m+j}\beta_j + t_{i,m+1}\beta_{m+1}), \quad c_{i,4m+2} = t_{i,m+1}.$$

From the definition of $\mathcal{N}_\mathcal{L}$ there are at least $K+1$ nonzero pairs, saying $(a_{i_j} | a_{N+i_j})$, $0 \le j \le K$, in the codeword $a$. Thus either

$(a_{i_j,1},\ldots,a_{i_j,m} | a_{N+i_j,1},\ldots,a_{N+i_j,m})$ or $(a_{i_j,m+1},\ldots,a_{i_j,2m} | a_{N+i_j,m+1},\ldots,a_{N+i_j,2m})$ is nonzero.

Without losing generalization, we suppose $(a_{i_j,1},\ldots,a_{i_j,m} | a_{N+i_j,1},\ldots,a_{N+i_j,m})$, $0 \le j \le K$ are nonzero. Then let us study vectors $(b_{i_j,1},\ldots,b_{i_j,2m+1} | c_{i_j,1},\ldots,c_{i_j,2m+1})$, $0 \le j \le K$ in the codeword $c$. From the definition of $b_{i,j}$ and $c_{i,j}$, it is easy to prove that among these $K+1$ nonzero vectors each vector may occur as many as $2^m$ times. This is to say, $c$ contains at least $(K+1)/2^m$ distinct nonzero binary $(4m+2)$-tuples. Now let $\omega$ be a primitive element of $GF(4)$ and let $c'$ be the quaternary vector

$$c' = (b_{0,1}+c_{0,1}\omega,\ldots,b_{0,4m+2}+c_{0,4m+2}\omega;\ldots;b_{N-1,1}+c_{N-1,1}\omega,\ldots,b_{N-1,4m+2}+c_{N-1,4m+2}\omega).$$

Then $c'$ contains at least $(K+1)/2^m$ distinct nonzero quaternary $(2m+1)$-tuples. From the choice of $K$,

$$\frac{K+1}{2^m} \ge \frac{1}{2}(1-\frac{2m+1}{m}R)\frac{2^{2m}-1}{2^m} \ge \frac{1}{2}(1-\frac{2m+1}{m}R)(2^m-1).$$

We can now apply Lemma 4 with $L = 2m+1$, $\delta = m/(4m+2)$, $\gamma = \frac{1}{2}[1-R(2m+1)/m]$, and deduce that

$$wt(c') \ge 2^m(2m+1)\frac{1}{2}(1-\frac{2m+1}{m}R)(2^m-1)(H_4^{-1}(\delta)-o(m)),$$

where the initial $2^m$ is because each $(2m+1)$-tuples occurs $2^m$ times (in the worst case). So

$$\frac{\text{distance}}{\text{length}} \geq \frac{2^{2m}-2^m}{2^{2m}-1}\frac{1}{4}(1-\frac{2m+1}{m}R)(H_4^{-1}(\frac{m}{4m+2})-o(m)) \doteq \frac{H_4^{-1}(1/4)}{4}(1-2R)$$

as $m \to \infty$. Q.E.D.

To sum up, if let $\delta$ and $R$ denote the lower bound to distance/length and rate of a family of quantum codes respectively as the length $n \to \infty$, we have found a family of asymptotically good concatenated quantum codes with $0 < R < \frac{1}{2}$ and $\delta = \frac{1}{4}H_4^{-1}(1/4)(1-2R)$. In fact in 1996 Calderbank and Shor [1] have proven the existence of good quantum codes. Then Calderbank et al. [5] proved the quantum Gilbert-Varshamov bound. But these proofs are not constructive. Later, Ashikhmin et al. [13] explicitly constructed asymptotically good quantum codes with $0 < \delta < \frac{1}{18}$, $R = 1-(2^{m-1}-1)^{-1}-\frac{10}{3}m\delta$ and Chen et al. [14] with $0 < \delta < \delta_t$ where $\delta_t = \frac{2}{3}(2^t-3)(2t+1)^{-1}(2^t-1)^{-1}$ for $t \geq 3$, $R = 3t(\delta_t-\delta)$. Then Matsumoto [15] improved the bound of Ashikhmin et al. [13] with $0 < \delta \leq (2m)^{-1}[\frac{1}{2}-(2^m-1)^{-1}]$, $R = 1-2(2^m-1)^{-1}-\frac{10}{3}m\delta$ for $m \geq 2$. But all these quantum codes are derived from good classical codes directly. Compared with above codes, although the performance of our code is not very excellent, this is the first time that good quantum codes are explicitly constructed from bad codes. Its occurrence supplies a gap in quantum coding theory.

＊Electronic address: lizhuo@xidian.edu.cn

† Electronic address: l_zhuo@21cn.com